\documentclass[aps,prb,twocolumn,citeautoscript,superscriptaddress]{revtex4-1}
\usepackage{amsmath,amssymb,mathrsfs,bm,setspace}
\usepackage{graphicx}
\usepackage{color}
\usepackage[colorlinks=true]{hyperref}
\hypersetup{
    bookmarks=true,         
    unicode=false,          
    pdftoolbar=true,        
    pdfmenubar=true,        
    pdffitwindow=false,     
    pdfstartview={FitH},    
    pdftitle={My title},    
    pdfauthor={Author},     
    pdfsubject={Subject},   
    pdfcreator={Creator},   
    pdfproducer={Producer}, 
    pdfkeywords={keyword1} {key2} {key3}, 
    pdfnewwindow=true,      
    colorlinks=true,       
    linkcolor=red,          
    citecolor=blue,        
    filecolor=magenta,      
    urlcolor=cyan           
}
\usepackage{bm}
\usepackage{latexsym}
\usepackage{placeins}
\def \r{{\mathbf{r}}}

\def \a{\alpha}
\def \1{\bf 1}
\def \2{\bf 2}
\def \A{{\mathbf{A}}}

\def \q{{\mathbf{q}}}

\def \tn{\textnormal}

\newcommand{\beq}{\begin{equation}}
\newcommand{\eeq}{\end{equation}}
\newcommand{\ba}{\begin{array}{ccc}}
\newcommand{\ea}{\end{array}}
\newcommand{\nn}{\nonumber \\}
\def\bea{\begin{eqnarray}}
\def\eea{\end{eqnarray}}

\begin{document}

\title{Diamagnetism and density wave order\\ in the pseudogap regime of YBa$_2$Cu$_3$O$_{6+x}$}
 \author{Lauren E. Hayward}
 \affiliation{Department of Physics and Astronomy, University of Waterloo, Ontario, N2L 3G1, Canada}
 \author{Andrew J. Achkar}
 \affiliation{Department of Physics and Astronomy, University of Waterloo, Ontario, N2L 3G1, Canada}
 \author{David G. Hawthorn}
 \affiliation{Department of Physics and Astronomy, University of Waterloo, Ontario, N2L 3G1, Canada}
 \author{Roger G. Melko}
 \affiliation{Department of Physics and Astronomy, University of Waterloo, Ontario, N2L 3G1, Canada}
 \affiliation{Perimeter Institute for Theoretical Physics, Waterloo, Ontario N2L 2Y5, Canada}
 \author{Subir Sachdev}
 \affiliation{Department of Physics, Harvard University, Cambridge, MA 02138, USA}
\affiliation{Perimeter Institute for Theoretical Physics, Waterloo, Ontario N2L 2Y5, Canada}
 \date{\today\\
}
\begin{abstract}
Clear experimental evidence of charge density wave correlations competing with superconducting
order in YBCO have thrust their relationship with the pseudogap regime into the spotlight.
To aid in characterizing the pseudogap regime,
we propose a dimensionless ratio of the 
diamagnetic susceptibility to the correlation length of the charge density wave correlations.
Using Monte Carlo simulations, we compute this ratio on the classical model of Hayward {\it et.~al.}
(Science {\bf 343}, 1336 (2014)),
which describes angular fluctuations of a multicomponent order, capturing both superconducting and density wave
correlations. We compare our results with available data on YBa$_2$Cu$_3$O$_{6+x}$, and propose experiments to clarify the value of this dimensionless ratio using existing samples and techniques.
\end{abstract}
\maketitle

\section{Introduction}
\label{sec:intro}

A fundamental characteristic of the pseudogap regime of the hole-doped cuprate superconductors
has been the presence of a large diamagnetic susceptibility over a wide range of temperatures above
the critical temperature for superconductivity\cite{ong00,ong0,ong,ybcomag}.
This behavior has been modeled in various theories of thermal fluctuations of the superconducting order
and its vortices\cite{larkin,vortex1,vortex2,vortex3}.

On the other hand, a seemingly different view of the pseudogap has emerged from recent X-ray scattering 
experiments\cite{keimer,chang,hawthorn,achkar2,comin1,comin2,neto13}.
In a regime of doping where the antiferromagnetic correlations are weak, these experiments observe
substantial charge density wave (CDW) correlations. The temperature and magnetic field 
dependence of these observations indicate that the CDW order competes with the superconducting (SC) order.

It is the purpose of the present paper to reconcile these distinct experimental probes of the pseudogap. 
First, one can measure the strength of the SC fluctuations by the diamagnetic susceptibility, $\chi_d = M/B$, where $M$ is the magnetization
per unit volume in the presence of a field $B$ applied perpendicular to the CuO$_2$ layers. Second, one can characterize the CDW correlations
by the value of their correlation length $\xi_{\rm cdw}$. From these quantities, which can be directly measured in experiments on 
the same sample in absolute units, we propose to form the following dimensionless ratio:
\beq
R(T) = 12 \pi s \left( \frac{\hbar}{2e} \right)^2 \frac{\chi_d}{k_{\rm B} T \, \xi_{\rm cdw}^2}. \label{eq:ratio}
\eeq
Here $e$ is the electron charge, $s$ is the interlayer spacing, $T$ is the absolute temperature, and the prefactor of $12 \pi$ is for numerical convenience. 

The utility of $R(T)$ extends to both experiment and theory. It is directly measurable from magnetic susceptibility 
and X-ray scattering experiments (preferably from the same sample), and offers a dimensionless measure
of the relative strength of the fluctuations of the order parameters for superconductivity and charge density waves.
Previous models of diamagnetism\cite{larkin,vortex1,vortex2,vortex3} have used phenomenological theories for superconducting fluctuations with a number of adjustable parameters. Our model has a similar effective theory for superconductivity, 
but the {\em same\/} parameters {\em also\/} determine the charge order fluctuations. By taking a dimensionless ratio, the theoretical
predictions become insensitive to the short-distance cutoff of the theory, and to the arbitrary scales used in defining the 
order parameters. Measurements and computations of $R(T)$ therefore offer a route to comparing our understanding
of the pseudogap to a more constrained theory.

\begin{center}
\begin{figure}[t]
        \begin{flushleft}
        \centerline{\includegraphics[height=0.4\textwidth]{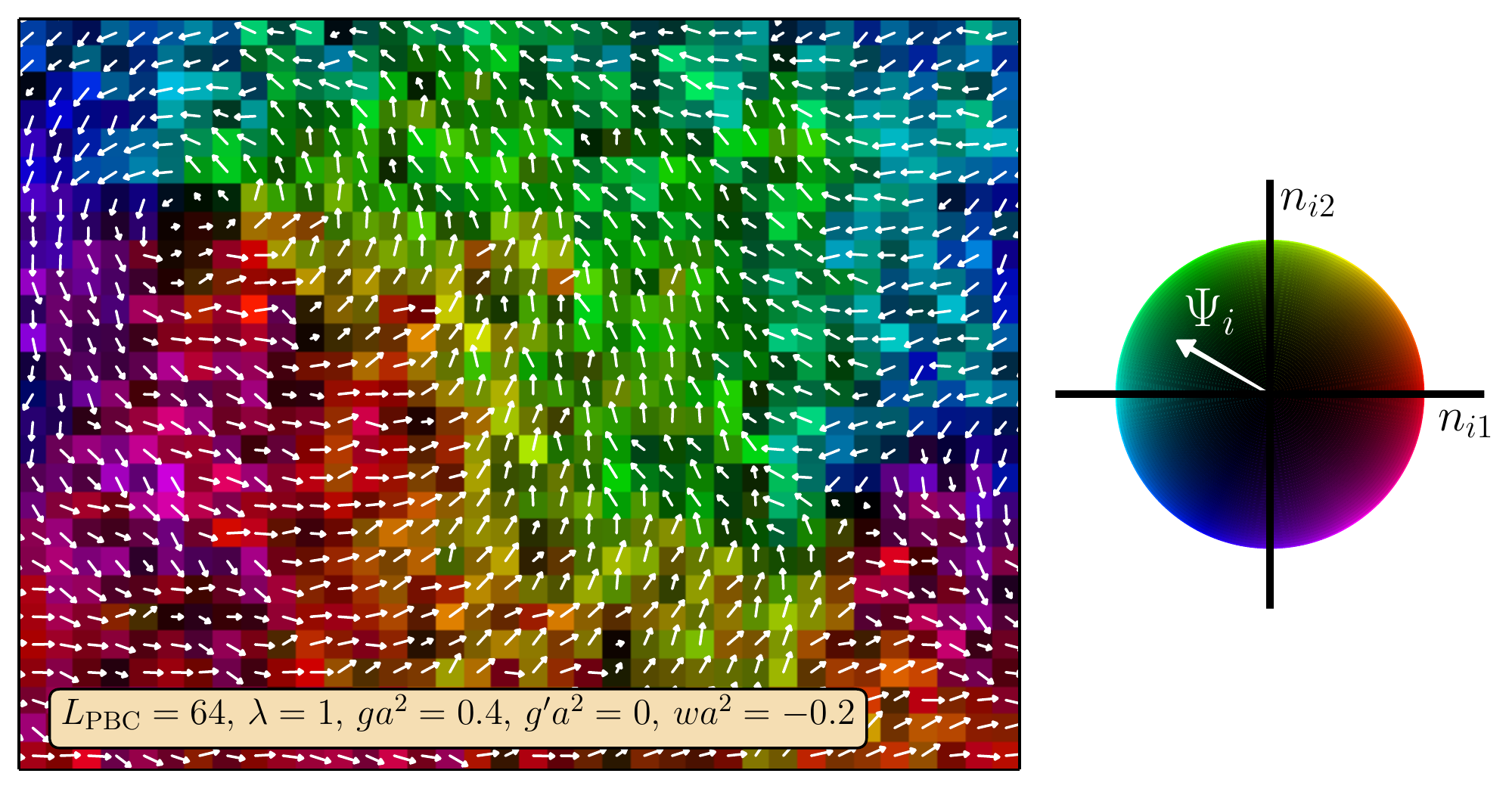}}

        \centerline{\includegraphics[height=0.4\textwidth]{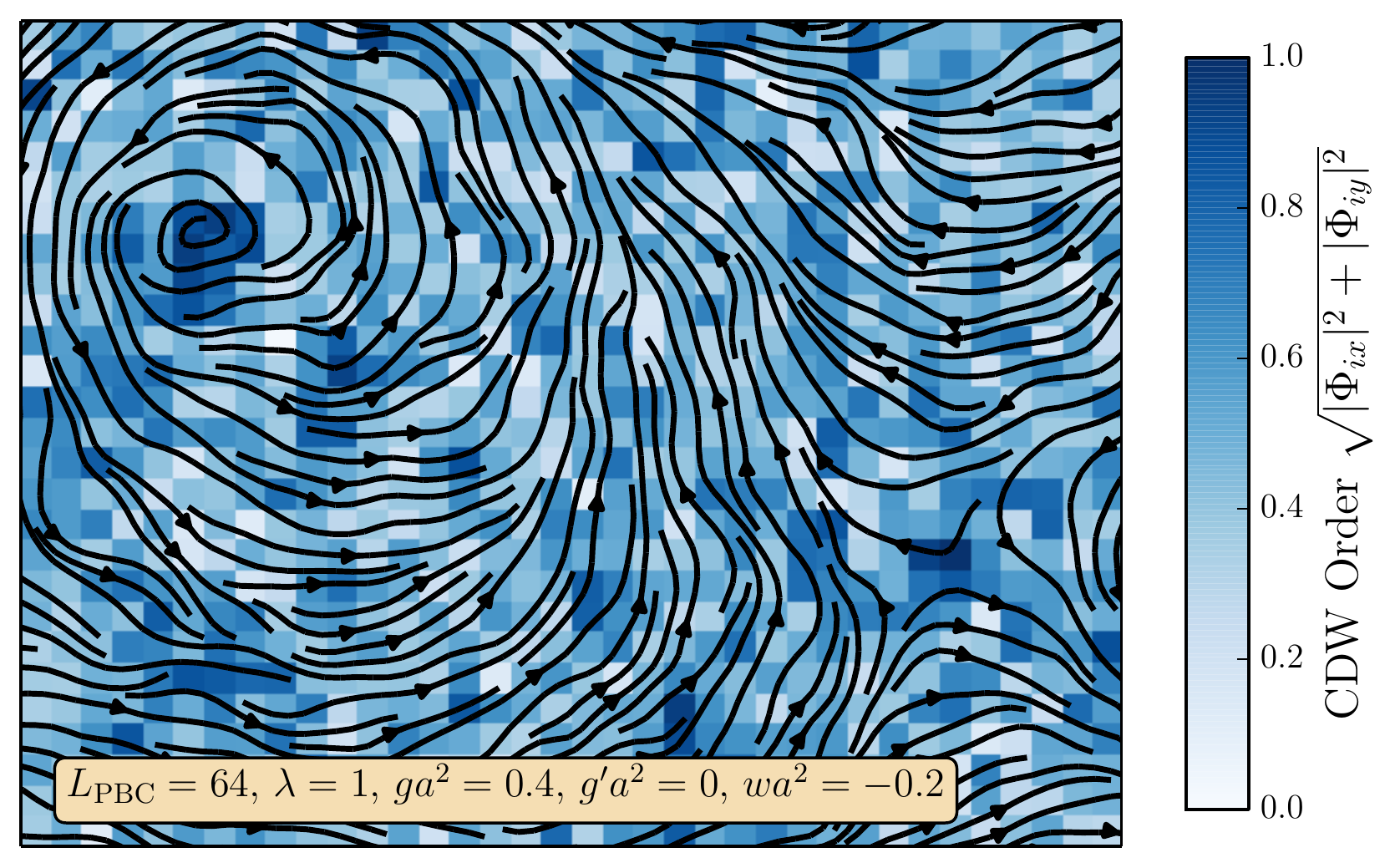}\hspace{0.87in}}
         \end{flushleft}
         \caption{Sample system configuration of the model of Ref.~\onlinecite{science} when $T/\rho_s = 0.18$ (below the Kosterlitz-Thouless transition). We show a 
         configuration of the variables $n_{i\alpha}$ on a subregion of a $64 \times 64$ lattice. We use a representative set of model parameters and employ two different visualization techniques. In the upper plot, we use shading to represent the strength of the CDW order parameter, with darker shading corresponding to stronger CDW magnitude $\sqrt{|\Phi_{ix}|^2 + |\Phi_{iy}|^2}$ on site $i$, and color to represent the orientation of the variables in the SC plane. The white arrows here correspond to the magnitude and orientation of the SC order $\Psi_i$. The lower plot uses color to represent the magnitude of the CDW order parameter and black arrows to illustrate the streamlines of $\Psi_i$. In both plots, a vortex-antivortex pair is visible in the upper left and lower right corners. The enhancement of CDW correlations in the vicinities of these vortices is also visible.
         }
         \label{fig:vortices}
\end{figure}
\end{center}

Experimentally, indications of a close relationship between superconductivity and the CDW order appeared already in the 
classic scanning tunneling microscopy observations of Hoffman {\it et al.\/} \cite{jenny}. These experiments observed
a CDW `halo' about each vortex in the superconducting order. In the pseudogap regime, the thermal fluctuations of vortex-antivortex
pairs are clearly the key to the diamagnetic response, as in the models of Refs.~\onlinecite{vortex1,vortex2,vortex3}; however these
works considered only `naked' vortices in the superconducting order, whose core did not possess any CDW correlations.  
Here we shall employ our recently proposed model of the pseudogap in Ref.~\onlinecite{science}, in which the superconducting
vortices are indeed linked to CDW correlations: this is captured by a snapshot from our Monte Carlo simulations in Fig.~\ref{fig:vortices}. 
Thus, in our model, the vortex fluctuations involved in the diamagnetic
response, are also directly tied to X-ray measurements of CDW correlations. We can, therefore, view $R(T)$ as a quantitative measure of the remarkable link
between the seemingly disparate superconducting and CDW orders.

Our model \cite{science} characterizes the CDW and SC fluctuations by a composite
order parameter with six real components, and focuses on classical and thermal fluctuations of this order along the angular directions
of the 6-dimensional space. In Ref.~\onlinecite{science}, the parameters of the

\onecolumngrid
\twocolumngrid

{\parindent0pt energy functional in the thermal partition function were constrained by 
comparing to X-ray data,\cite{science} which reflected the strong coupling between the SC and CDW order parameters.
Here, we present our results for $R(T)$ for a similar range of parameters.
We compare these computations with the available data on YBa$_2$Cu$_3$O$_{6+x}$, although the present diamagnetic susceptibility and X-ray data are not on the same sample. 
Despite these caveats, we show below that 
the absolute theoretical and experimental values of $R(T)$ are quite close to each other, and their $T$-dependencies are very similar.
We hope that our theoretical calculations will motivate experimental measurements of $R(T)$ on a single sample in the near future.}

\FloatBarrier
\section{Theoretical Model and Measurements}
\label{sec:model}

The model of Ref.~\onlinecite{science} describes the pseudogap using a 
non-linear sigma model (NL$\sigma$M) with classical variables $n_{i\a}$ ($\a=1 \ldots 6$) on sites $i$ of a square lattice, with the constraint $\sum_\a n_{i \a}^2 = 1$ on every site $i$.  These variables describe the SC order $\Psi$, the CDW order $\Phi_x$ along
the $x$ direction, and the CDW order $\Phi_y$ along the $y$ direction via
\bea 
\Psi &=& n_1+in_2 \nn
\Phi_x &=& n_3+in_4 \nn
 \Phi_y &=& n_5+in_6. 
\eea
The partition function, $\mathcal{Z}$, of the NL$\sigma$M is given by
\beq
\mathcal{Z}
=\prod_i\bigg[\int dn_{n_{i\a}}~\delta\bigg(\sum_{\alpha=1}^6n_{i\alpha}^2-1 \bigg)\bigg]~\tn{exp}\bigg(- \frac{H_1+H_2}{k_B T} \bigg),\label{eq:partFunc}
\eeq
where
\bea
H_1&=& \frac{\rho_s}{2} \sum_{\langle ij\rangle}\bigg[\sum_{\a=1}^2 (n_{i\a} - n_{j\a})^2 + \lambda\sum_{\a=3}^6
(n_{i\a} - n_{j\a})^2 \bigg], \nn
{H_2} &=& {\rho_s a^2}\sum_i \Biggl[ \frac{g}{2} \sum_{\a=3}^6 n_{ia}^2 
+ \frac{g'}{2} \left( \sum_{\a=3}^6 n_{ia}^2 \right)^2 \nn
&~&~~~~~~~~~~+ \frac{w}{2}\left[(n_{i3}^2+n_{i4}^2)^2+(n_{i5}^2+n_{i6}^2)^2\right] \Biggr]. \label{eq:H2}
\eea
The couplings $\rho_s$ and $\rho_s \lambda$ are the helicity moduli for spatial variations of the SC and CDW orders, 
and $g$ measures the anisotropy in the energy between the CDW and SC directions. 
We also allow here for a quartic anisotropy $g'$, which was not included in Ref.~\onlinecite{science}, in order to obtain a wider range of physical properties. 
The continuum theory is discretized on a lattice of spacing $a$. The lattice spacing $a$ will cancel out of our computations for the value of $R(T)$.

The ground state of $\mathcal{Z}$ at $T=0$ is easily determined: the optimal state is spatially uniform with $H_1 = 0$,
and we have to minimize $H_2$. Let us first take $w < 0$ so that the CDW is stripe-like {\em i.e.\/} only one of $\Phi_x$ or $\Phi_y$ 
is non-zero. Then, without loss of generality we can take
\beq
n_\alpha = ( \cos \theta, 0, \sin\theta, 0, 0,0),
\eeq
where $\theta=0$ corresponds to SC order, $\theta = \pi/2$ corresponds to CDW order, and anything in between is co-existence, namely SC+CDW.
Then, per site,
\beq
\frac{H_2}{\rho_s a^2}  = \frac{g}{2} \sin^2 \theta + \frac{w + g'}{2} \sin^4 \theta.
\eeq
Minimizing this function gives the parameter-dependent ground state
\bea
{\rm SC} \quad & : & g > 0,  \quad w+g' +g > 0 \nn
{\rm CDW} \quad &: & g > 0, \quad  w+g' +g < 0 \nn
{\rm CDW} \quad &:& g < 0, \quad w + g' + g/2 < 0 \nn
{\rm SC+CDW} \quad &:& g < 0, \quad w + g' + g/2 > 0. \label{phases1}
\eea \\

Next, we consider $w > 0$ so that the CDW is `checkerboard'. 
Then, without loss of generality we can take
\beq
n_\alpha = \left( \cos \theta, 0, \frac{\sin\theta}{\sqrt{2}}, 0,\frac{\sin\theta}{\sqrt{2}},0 \right),
\eeq
and then
\beq
\frac{H_2}{\rho_s a^2} = \frac{g}{2} \sin^2 \theta + \frac{w/2 + g'}{2} \sin^4 \theta \, ,
\eeq
so that the minima are as in Eq.~\eqref{phases1}, but with $w \rightarrow w/2$.

At finite temperature, an exact solution to the model is unavailable, and we turn to a numerical solution based on Monte Carlo
simulations as described in the next section.
There, we  focus on using the dimensionless ratio $R(T)$ to characterize the various phases of the model, for which one requires 
calculation of the susceptibility and correlation length.
We now describe the general method for computing the magnetization, $M$, of $\mathcal{Z}$ in the presence of an applied magnetic field.
We access $M$ on a $L \times L$ lattice with open boundary conditions using the method of Ref.~\onlinecite{stroud}, which allows us to avoid introducing flux quantization, which therefore means that we can consider the effect of arbitrarily small magnetic fields $B$
(in contrast, for example, to the cylindrical boundary conditions used in Refs.~\onlinecite{vortex2,vortex3}).
The expression for the magnetization can be computed by introducing an external vector potential $\A$.  We assume that there is only an orbital coupling to the superconducting order, $\Psi_j=n_{j1}+in_{j2}$. The contribution to the kinetic part from this coupling between $\Psi$ and $\A$ is then written as
\bea
H_\Psi &=& H_1 \bigg|_\Psi \\
&=& \frac{\rho_s}{2}\sum_{\langle ij\rangle}\bigg( |\Psi_i|^2 + |\Psi_j|^2 - \Psi_i^*\Psi_je^{iA_{ij}}-\Psi_j^*\Psi_ie^{-iA_{ij}}\bigg), \nonumber
\eea
where
\beq
A_{ij}=\frac{2e}{\hbar}\int_{\r_i}^{\r_j}d\r.\A.
\eeq
We will henceforth drop factors of $2e$ and $\hbar$.
We take $L$ even and
place the origin of co-ordinates at the center of the central plaquette. Thus, the sites are at
\beq
\r_i \equiv (x_i, y_i) = \left( i_x - \frac{L+1}{2} , i_y - \frac{L+1}{2} \right)a ,
\eeq 
with $i_{x,y} = 1 \ldots L$.
It is now convenient to label the site-dependence of the vector potential as $A_{iu}$, where $\mathbf{u}$ extends over $\pm \hat{x} a$ and $\pm \hat{y}a$
for bulk sites, and a smaller range for sites on the edge. Then we can write $H_\Psi$ as 
\bea
H_\Psi &=& \frac{\rho_s}{2} \sum_i Z_i |\Psi_i|^2 - \frac{\rho_s}{2} \sum_{i,u} \Psi_{i}^\ast \Psi_{i+u} e^{i A_{iu}} \nn
&=& \frac{\rho_s}{2} \sum_i Z_i |\Psi_i|^2 - \frac{\rho_s}{2} \sum_{i,u} \Psi_{i+u}^\ast \Psi_{i} e^{-i A_{iu}},
\eea
where $Z_i$ is the co-ordination number for site $i$. Thus the sum over $u$ always extends over $Z_i$ values.
Note that $A_{i+u,-u} = - A_{iu}$. The current flowing along link $iu$ is then
\beq
{\bf J}_{iu} = \mathbf{u} \frac{\rho_s}{2} \left(i \Psi_i^\ast \Psi_{i+u} e^{i A_{iu}} + \mbox{c.c.} \right).
\eeq
Finally, following Ref.~\onlinecite{stroud}, we can write the total magnetic moment divided by the volume as
\bea
M &=& \frac{1}{4L^2 a^2 s} \sum_{i,u} \r_i \times \left\langle \mathbf{J}_{iu} \right\rangle \nn
&=& \frac{\rho_s}{4L^2 a^2 s} \sum_{i,u} \epsilon_{\alpha\beta} r_{i \alpha} u_{\beta} \left\langle i \Psi_i^\ast \Psi_{i+u} e^{i A_{iu}} \right\rangle.
\eea
We apply a uniform magnetic field $B$ perpendicular to the plane, and choose the vector potential in the circular gauge
\beq
A_{iu} =  \frac{B}{2} \epsilon_{\alpha\beta} r_{i\alpha} u_{\beta}.
\eeq
Now we expand $M$ to first order in $B$ and obtain
\bea
\chi_d \equiv \frac{M}{B} &=& - \frac{\rho_s}{8 L^2 a^2 s} \sum_{i,u}  \left(\epsilon_{\alpha\beta} r_{i \alpha} u_{\beta}\right)^2 \left\langle \Psi_i^\ast \Psi_{i+u} \right\rangle_0  \nn
&+& \frac{\rho_s^2}{16T L^2 a^2 s} \sum_{i,u} \sum_{j,u'}  \left(\epsilon_{\alpha\beta} r_{i \alpha} u_{\beta}\right)\left(\epsilon_{\gamma\delta} r_{j \gamma} u'_{\delta}\right) \nn
&~&~~~~~~~~~~~~~~~~~~\times \left\langle \Psi_i^\ast \Psi_{i+u} \Psi_{j+u'}^\ast 
\Psi_j \right\rangle_0, 
\label{eq:diamag_Phi}
\eea
where the subscript $0$ indicates that the averages can be evaluated in zero field under $\mathcal{Z}$.
Note that the expression (\ref{eq:diamag_Phi}) is proportional to $a^2$ (after accounting for the powers of $a$ in
${\bf r}_i$ and ${\bf u}_i$); this factor of $a^2$ will cancel with that in $\xi_{\rm cdw}$ when we compute the ratio 
$R(T)$.

As described in the next section, we use this expression for $\chi_d$ to calculate the linear-order diamagnetic susceptibility of the model of Eq.~(\ref{eq:partFunc}).
However, before proceeding with the full NL$\sigma$M, we carefully
benchmark our expression for $M/B$ by calculating it for a simple Gaussian model, where exact analytical results are available.
As described in Appendix~\ref{app:gauss}, we find excellent agreement between our Monte Carlo results and 
Feynman diagram computations for this Gaussian model.  

\FloatBarrier
\section{Monte Carlo Simulation}
\label{sec:simulation}

By using classical Monte Carlo techniques, one can solve for thermodynamic properties of the model of Ref.~\onlinecite{science}, i.e., Eq.~\eqref{eq:partFunc}, on a finite-size lattice.
These techniques involve importance-sampling of configurations of the classical variables $n_{i \alpha}$ according to a Boltzmann probability distribution.
In order to generate independent configurations weighted by the partition function $\mathcal{Z}$, we use a combination of local\cite{metropolis, hastings, newman} and non-local\cite{wolff, newman} sampling techniques. 
Both of these sampling techniques require a method for generating a random point (corresponding to the coordinates $n_{i\alpha}$) on a hypersphere in 6-dimensional space. 
In order to generate such a random point, we choose each of the coordinates $n_{i\a}$ from a normal distribution and then project the resulting point onto the surface of the hypersphere\cite{marsaglia}.

The non-local sampling consists of a modified Wolff cluster update, where we add to the standard Wolff algorithm\cite{wolff} a cluster acceptance probability to account for the onsite energy terms in $H_2$ in Eq.~\eqref{eq:H2}. 
As expected, the non-local sampling provides notable efficiency gain (which becomes more significant as the temperature decreases) and also helps to prevent the ergodicity loss that can occur at low temperatures. 
We note that the non-local cluster sampling described above is not possible when $\lambda \neq 1$ due to the anistropic coupling that results between the hyperplanes corresponding to $\Psi$ and $\Phi_\mu$ ($\mu = x, y$). 
However, at moderate temperatures, it is still possible to obtain reasonable results with $\lambda \neq 1$ using only local sampling.

\begin{figure}
\includegraphics[width=0.45\textwidth]{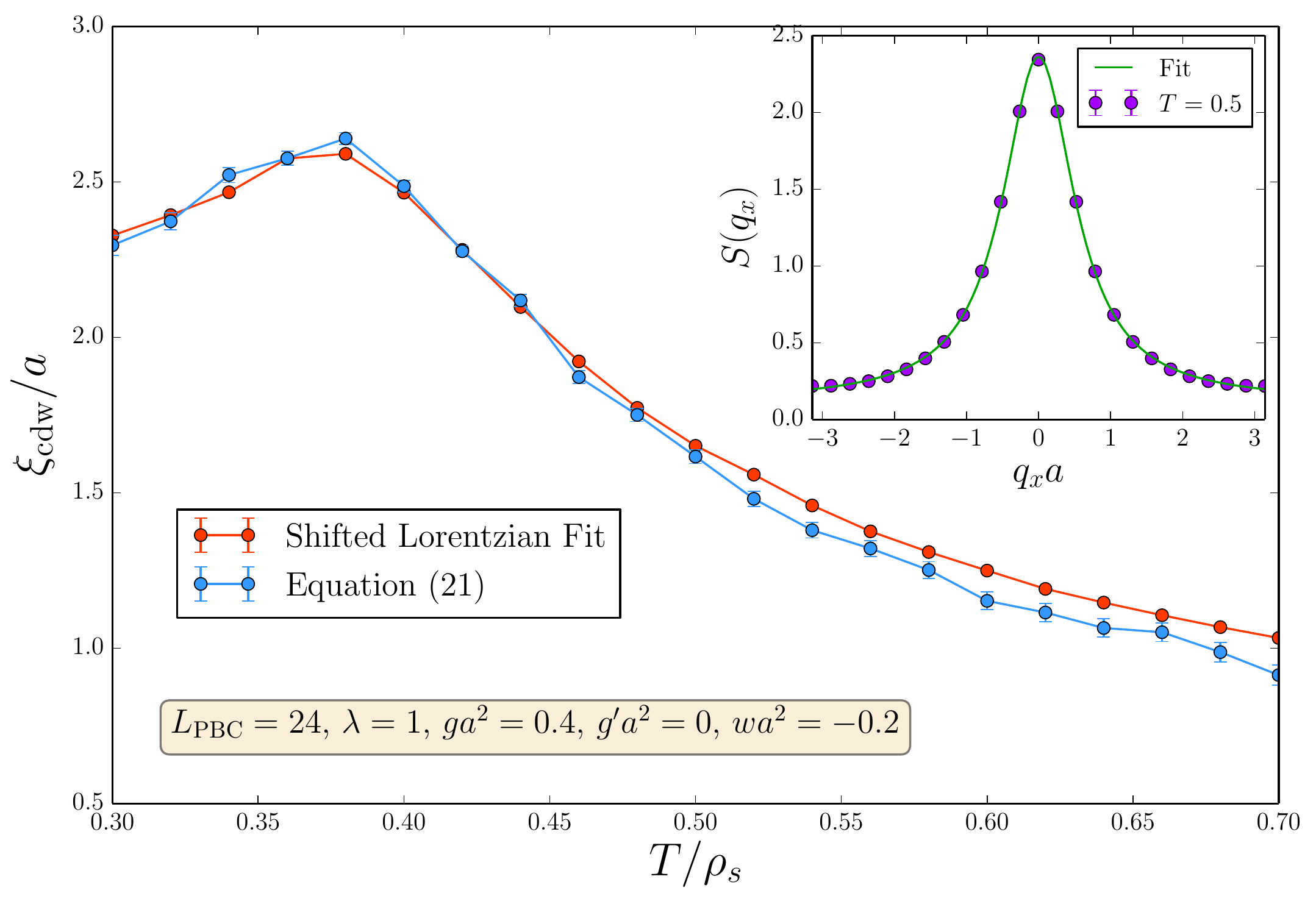} 
\caption[]{The CDW correlation length $\xi_{\rm cdw}$ extracted from Monte Carlo simulations. We illustrate two different methods for extracting the correlation length for a representative set of parameters.  The first method fits the structure factor $S(q_x)$ to the shifted Lorentzian function of Eq.~(\ref{LorentzShift}) for each $T$, as illustrated in the inset for $T=0.5$. Error bars come from the covariance matrix of the least-squares fit. The second method is to calculate $\xi_{\rm cdw}$ from Eq.~\eqref{eq:corrLen_Sandvik}, with error bars corresponding to the statistical Monte Carlo error.  
}
\label{fig:corrLenVsT}
\end{figure}

Using such updates, a Monte Carlo procedure is capable of calculating all standard estimators, such as the energy and magnetization.
In order to compute $R(T)$ in Eq.~\eqref{eq:ratio}, one needs to access two specific quantities, 
namely the linear-order diamagnetic susceptibility $M/B$, 
and the correlation length $\xi_{\rm cdw}$ of CDW correlations. 
We begin by describing our procedure for the latter.
We perform Monte Carlo simulations to compute the CDW correlation function 
\beq
C_{\Phi_x}(\r_i - \r_j) = \left\langle \sum_{\alpha=3}^4 n_{i \a}n_{j \a} \right\rangle
\eeq
on an $L \times L$ lattice,  with periodic boundary conditions (in order to minimize potential edge effects). 
We then calculate the structure factor 
\beq
S_{\Phi_x}(\q) = \sum_{\r} C_{\Phi_x}(\r) \cos(\q \cdot \r)
\label{eq:structFact}
\eeq 
with $\q = q_x \hat{x}$, and compare various methods for extracting $\xi_{\rm cdw}$.
The first such method involves a least-squares fit of $S_{\Phi_x}(q_x)$ to a shifted Lorentzian function, 
\beq
A \left( q_x^2 + 1/\xi^2 \right)^{-1} + c. \label{LorentzShift}
\eeq 
We add the shift $c$ to the Lorentzian fitting function to account for the effects of the missing short-wavelength degrees of freedom, which are significant here since $\xi_{\rm cdw}$ is of the order of the lattice spacing $a$. 
The second method, as described in Ref.~\onlinecite{sandvik}, is obtained by assuming the Ornstein-Zernike form for the correlation function and subsequently calculating $\xi_{\rm cdw}$ from 
\beq
\xi_{\rm cdw} = \frac{L}{2\pi}\sqrt{\left( \frac{8d}{(1+d)(3+d)} \right) \left(\frac{S_{\Phi_x}(0)}{ S_{\Phi_x}\left(\frac{2\pi}{La} \hat{x} \right) } - 1\right) }. \label{eq:corrLen_Sandvik}
\eeq
Results for $\xi_{\rm cdw}$ vs. $T$ are shown in Figure~\ref{fig:corrLenVsT} for $L=24$. 
Note that careful finite-size scaling analysis concludes that the data for $\xi_{\rm cdw}$ extracted through these procedures is
converged by lattice size $L=24$ for the model parameters studied in this paper.

\begin{figure}[h]
	\includegraphics[width=0.45\textwidth]{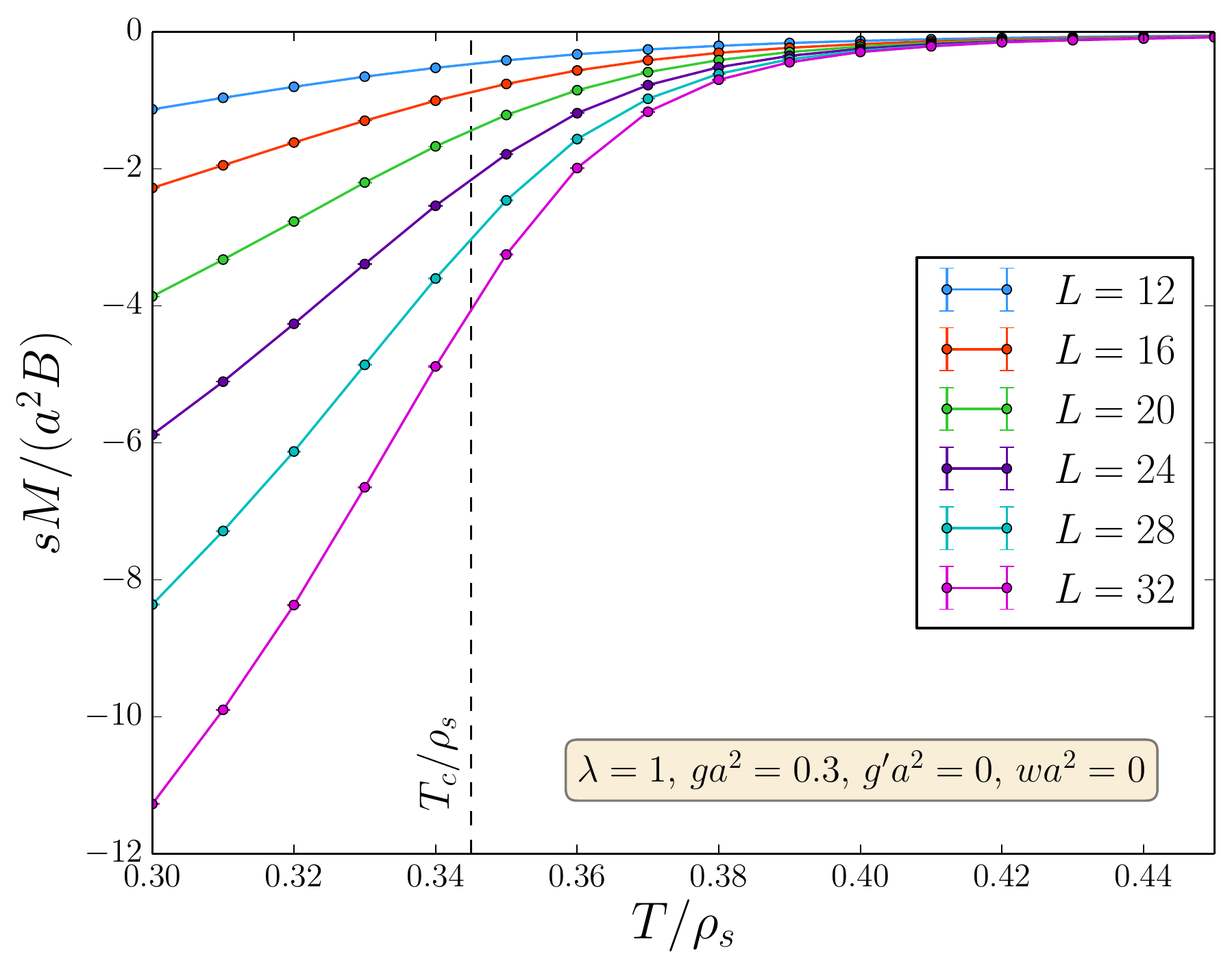} 
	\caption{ Monte Carlo calculations of the linear diamagnetic susceptibility $M/B$ . We plot $sM/(a^2 B)$ for a representative set of parameters and various $L$, with $M/B$ calculated from Eq.~\eqref{eq:diamag_MC} in a Monte Carlo simulation on a system with open boundary conditions. The dashed line is the location of the
	Kosterlitz-Thouless transition.}
         \label{fig:diamagVsT}
\end{figure}

Next, we also 
use Monte Carlo methods to calculate the linear-order diamagnetic susceptibility, $M/B$. For these calculations, we write Eq.~\eqref{eq:diamag_Phi} in terms of the coordinates $n_{i 1}$ and $n_{i 2}$ as 
\bea
\chi_d \equiv \frac{M}{B} &=& -\frac{\rho_s}{4 L^2 s} \sum_i \sum_{u= +\hat{x}a, +\hat{y}a} \left( \epsilon_{\alpha \beta} r_{i \alpha} u_{\beta} \right)^2 \nn
&~&~~~~~~~~~~~~~~~~\times \left\langle n_{i1} n_{i+u,1} + n_{i2} n_{i+u,2} \right\rangle_0 \nn
& +& \frac{\rho_s^2}{4 T L^2 s} \Biggl\langle \biggl[ \sum_i \sum_{u= +\hat{x}a, +\hat{y}a} \left( \epsilon_{\alpha \beta} r_{i \alpha} u_{\beta} \right) \nn
&~&~~~~~~~~\times
 \left( n_{i1} n_{i+u,2} - n_{i2} n_{i+u,1} \right) \biggr]^2 \Biggr\rangle_0. \label{eq:diamag_MC}
\eea
These simulations are performed separately from those for calculating $\xi_{\rm cdw}$ since, as described in Section~\ref{sec:model}, our method for calculating $M/B$ requires a lattice with open boundary conditions. Results for the diamagnetic susceptibility are shown for various lattice lengths $L$ for a given set of parameters $\lambda$, $g$, $g'$ and $w$ in Fig.~\ref{fig:diamagVsT}.
For $T>T_c$, $\chi_d = M/B$ converges well to a limiting value as $L \rightarrow \infty$. We discuss the situation below the Kosterlitz-Thouless
transition in Appendix~\ref{app:super}: there we show that $s \chi_d =  - 0.0351 \rho_s^R (La)^2$, where $\rho_s^R$ is the renormalized stiffness;
the divergence as $L \rightarrow \infty$ is a manifestation of the Meissner effect.

\FloatBarrier
\section{Results}

\begin{figure}
	\includegraphics[width=0.45\textwidth]{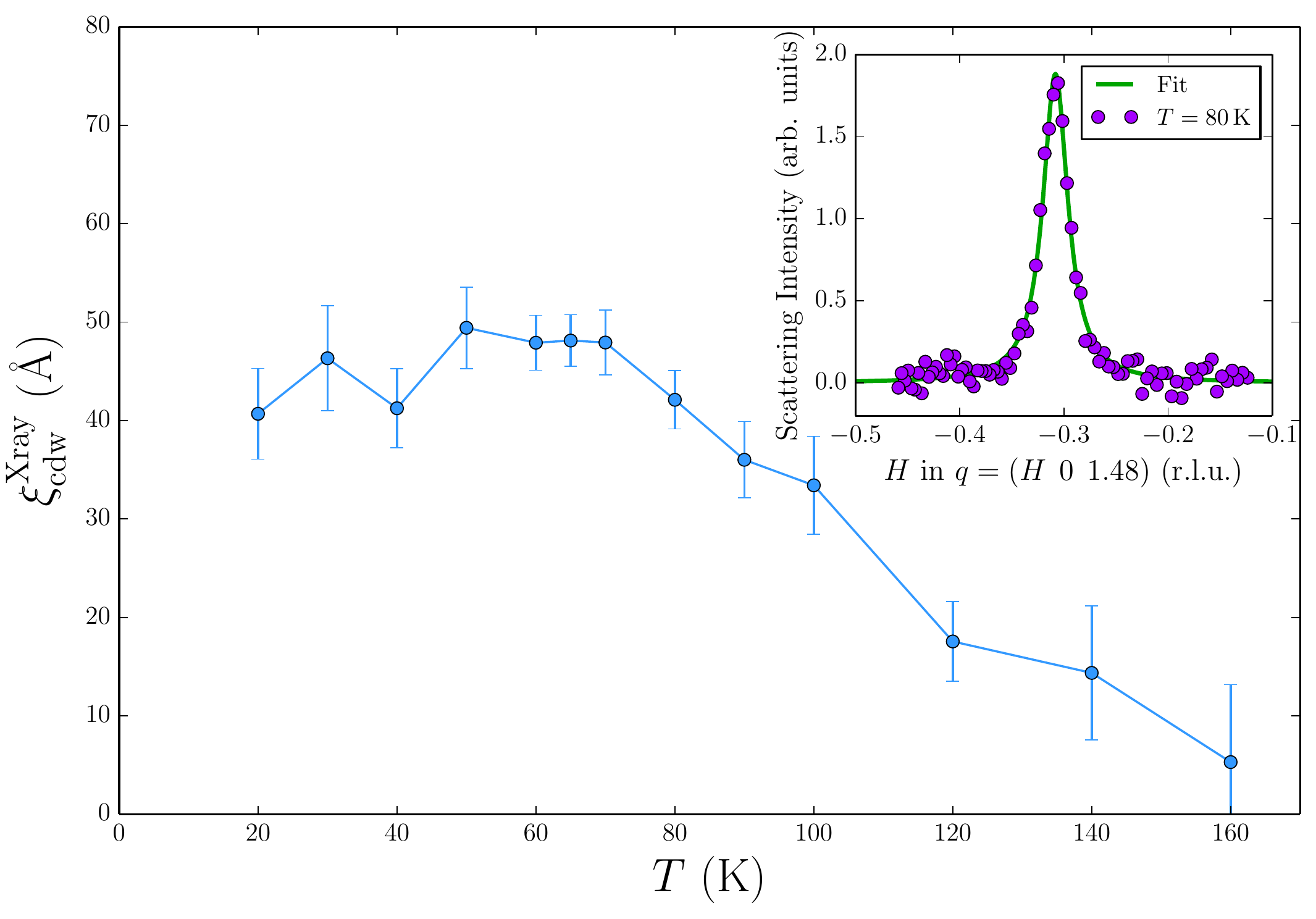} 
	\caption{The CDW correlation length $\xi_{\rm cdw}^{\rm Xray}$ extracted from X-ray scattering experiments\cite{achkar2} on YBa$_2$Cu$_3$O$_{6.67}$ as a function of $T$. $\xi_\text{cdw}^\text{Xray}(T)$ is extracted from Lorentzian fits to background-subtracted X-ray scattering data. The inset illustrates this fit for data at $T=80\,$K. The error bars reflect the statistical uncertainty in the fit and additional uncertainty due to the background subtraction. }
         \label{fig:corrLenVsT_Exp}
\end{figure}

Using our Monte Carlo calculations of the CDW correlation length and linear diamagnetic susceptibility in the model described above, we are
ready to calculate the dimensionless ratio $R(T)$ to compare to experiment.
In order to calculate the experimental quantity, 
we write the measurements of the diamagnetic susceptibility in the form, following Ref.~\onlinecite{ybcomag},
\beq
M (T) \equiv - \left( \frac{2e}{\hbar} \right)^2 \frac{T B}{ 12 \pi s }  \xi_{ab}^2 (T) \label{defxi}
\eeq
where we take Eq.~\eqref{defxi} as the {\it definition\/} of the length $\xi_{ab} (T)$, which is determined from torque magnetometry experiments on underdoped YBa$_2$Cu$_3$O$_{6.5}$ with $T_c=57\,$K \cite{cooper}.
Then the experimental value of the ratio $R(T)$ is simply
\beq
R(T) = - \left( \frac{\xi_{ab} (T)}{\xi_{\rm cdw}^{\rm Xray}(T)} \right)^2 \label{Rexp}
\eeq
where $\xi_{\rm cdw}^{\rm Xray}(T)$ is the correlation length of the charge order determined from X-ray scattering experiments on oxygen-disordered YBa$_2$Cu$_3$O$_{6.67}$ with $T_c=65.5\,$K \cite{achkar2}. To extract this correlation length, we first subtract the X-ray fluorescence background using a measurement at 160$\,$K and then fit the resulting profile using a Lorentzian function. Results for $\xi_{\rm cdw}^{\rm Xray}(T)$ vs. $T$, as well as this fitting procedure, are illustrated in Figure~\ref{fig:corrLenVsT_Exp}.

\begin{figure}[h]
        \centering
        	\includegraphics[width=0.4\textwidth]{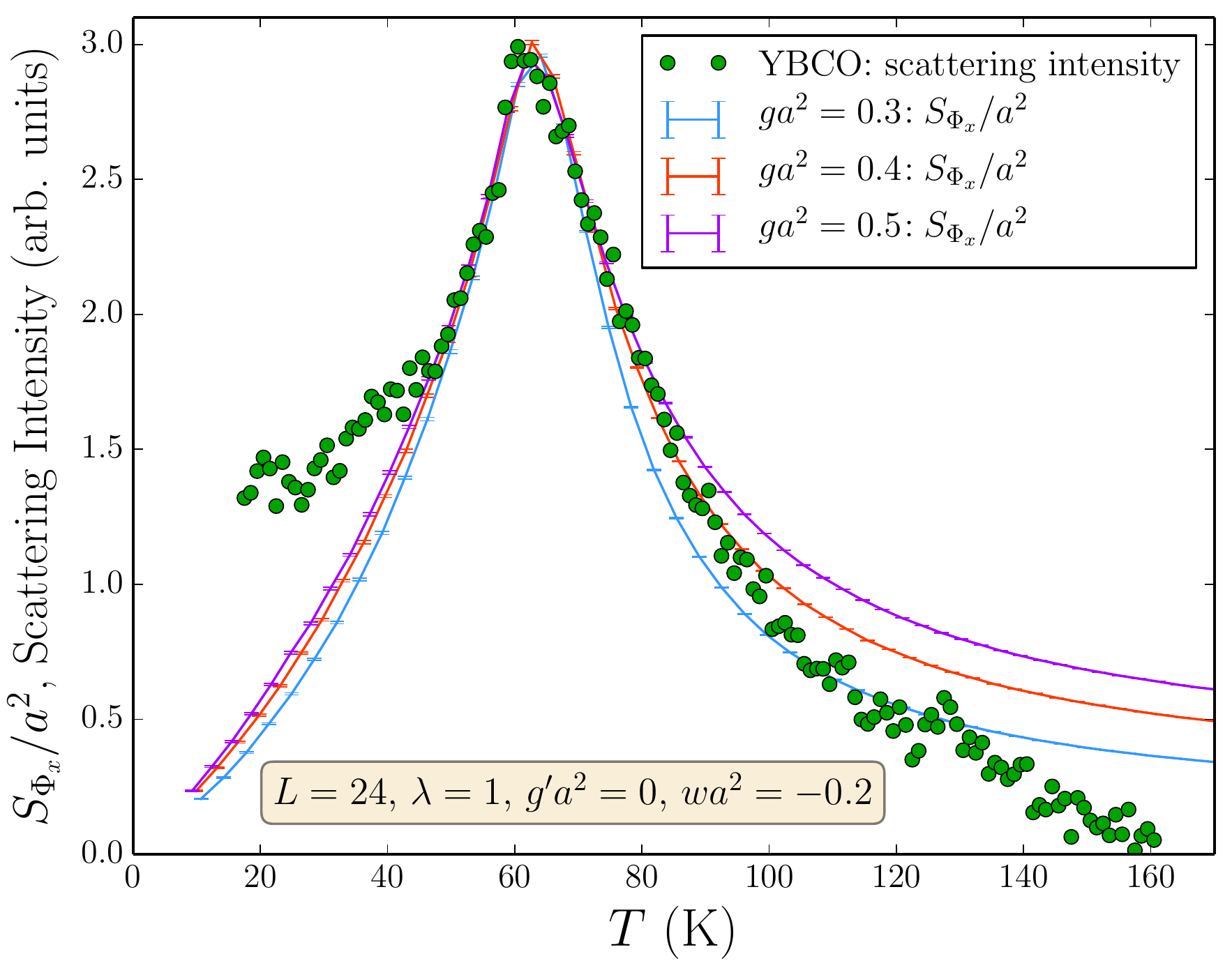} 
        
        	\includegraphics[width=0.4\textwidth]{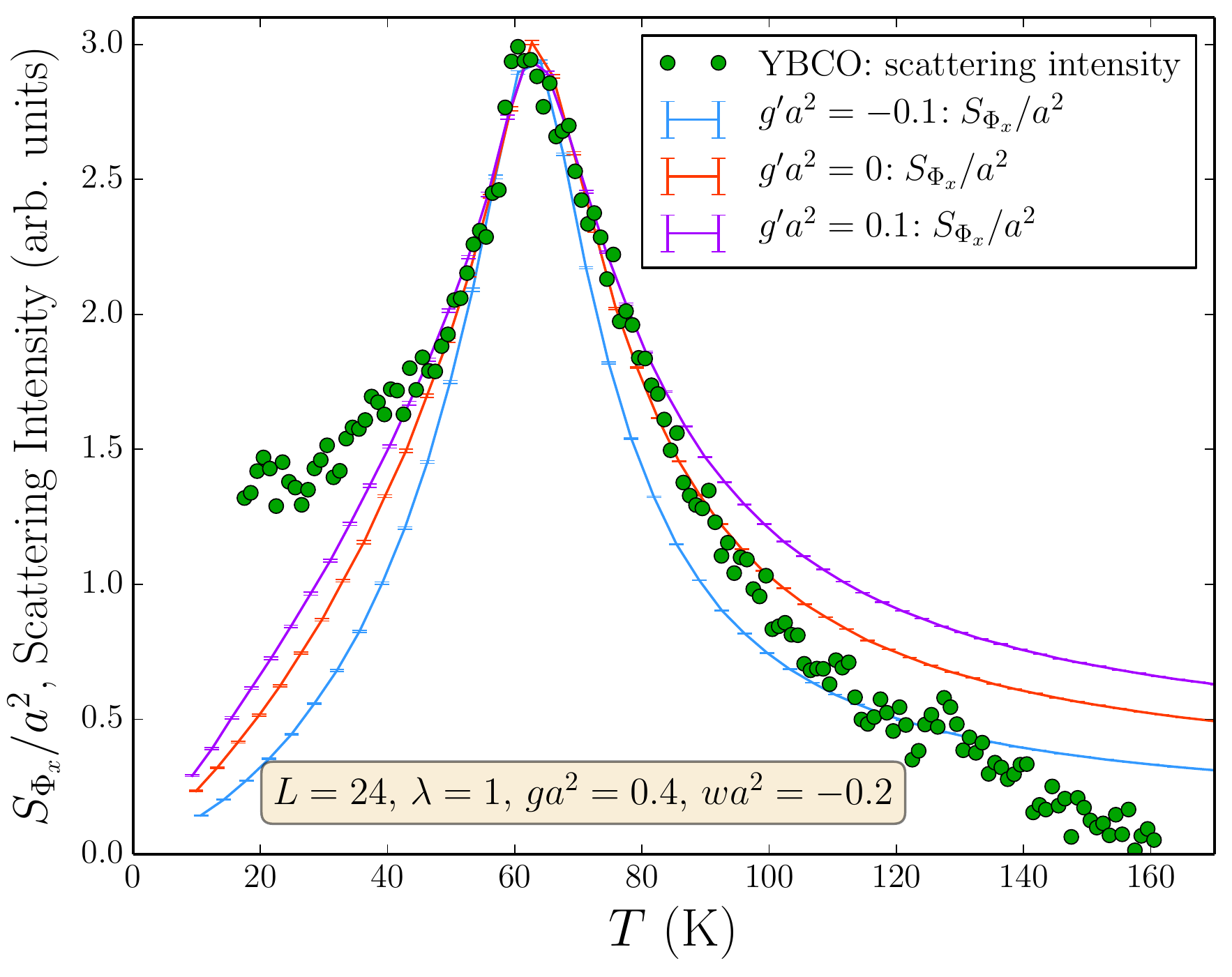}
        
        	\includegraphics[width=0.4\textwidth]{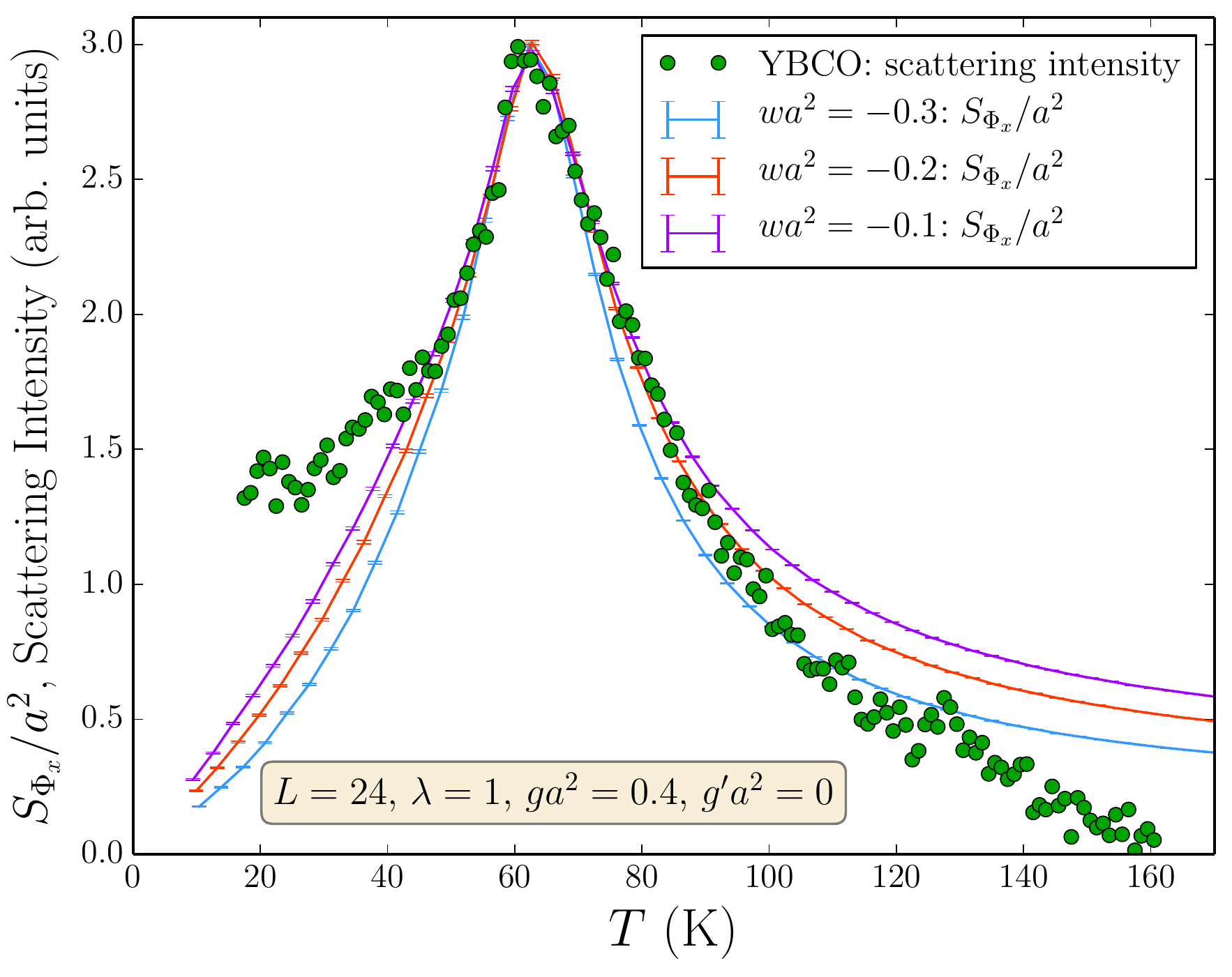}
         \caption{Comparison of $S_{\Phi_x}$ calculated in Monte Carlo simulations to X-ray data from CDW scattering experiments\cite{achkar2} on YBa$_2$Cu$_3$O$_{6.67}$. We illustrate results that demonstrate the effects of varying $g$ (top), $g'$ (middle) and $w$ (bottom) in our model. Note that, following the procedure of Ref.~\onlinecite{science}, there are two fitting parameters for each set of parameters $\lambda$, $ga^2$, $g'a^2$ and $wa^2$: the value of $\rho_s$ as well as the vertical scaling factor were both adjusted to make the Monte Carlo and experimental curves match in the vicinity of the peak. 
         }
         \label{fig:SPhiVsT}
\end{figure}

In order to compare these results to our Monte Carlo calculations of the dimensionless ratio in Eq.~\eqref{eq:ratio}, 
we must first determine the value of $\rho_s$ for each set of parameters $\lambda$, $ga^2$, $g'a^2$ and $wa^2$ in our model.
To do this, we use the prodecure of Ref.~\onlinecite{science} to compute the structure factor in Eq~\eqref{eq:structFact} with $\q = 0$.
We compare our results for $S_{\Phi_x} (\q = 0)$ with CDW scattering intensities from X-ray scattering experiments and determine $\rho_s$ (as well as the vertical scaling factor for the Monte Carlo data) by requiring that the curves match in the vicinity of the peak. 
This procedure is illustrated in Fig.~\ref{fig:SPhiVsT}.

\begin{figure}[h]
        \centering
        	\includegraphics[width=0.4\textwidth]{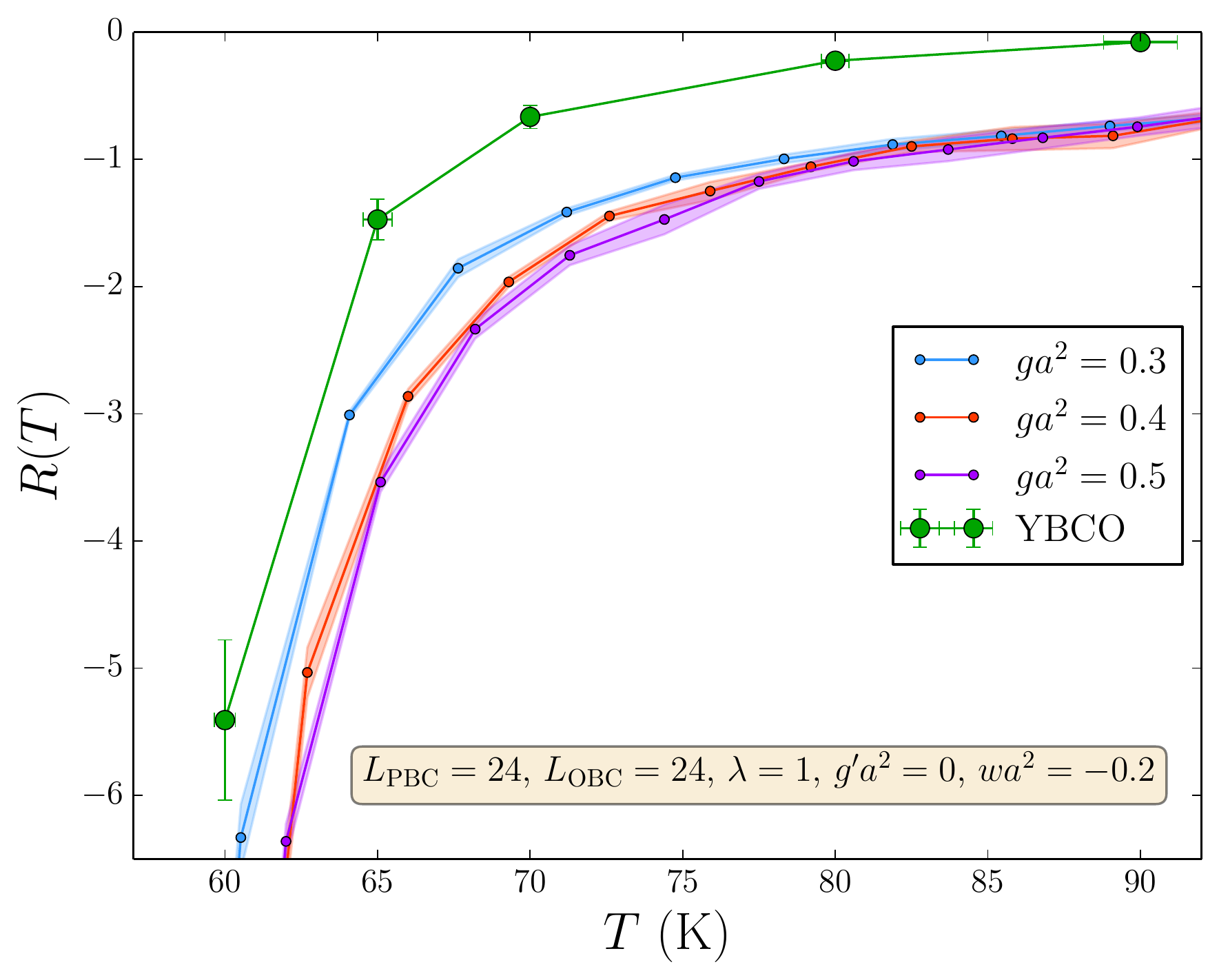} 
        
        	\includegraphics[width=0.4\textwidth]{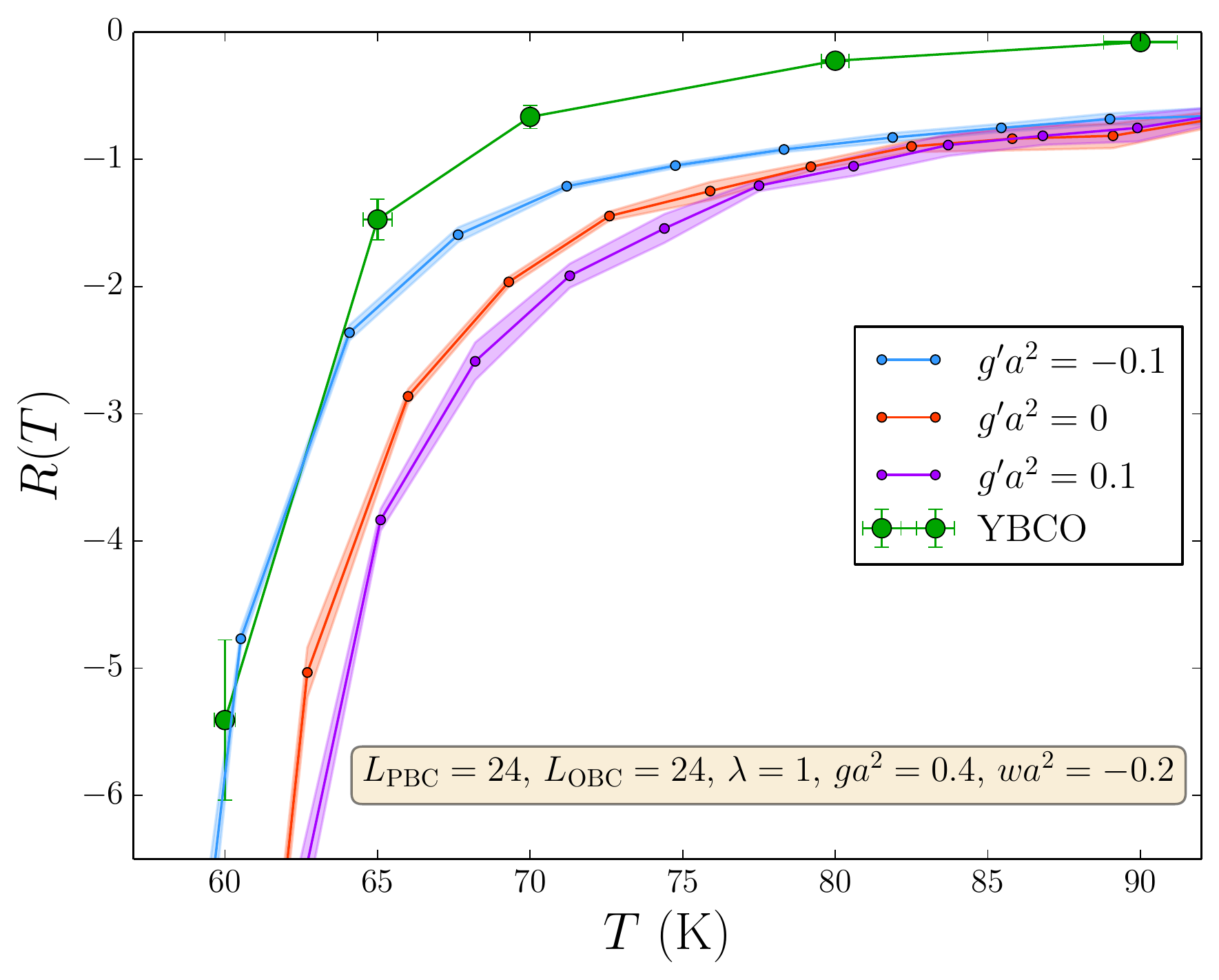}
        
        	\includegraphics[width=0.4\textwidth]{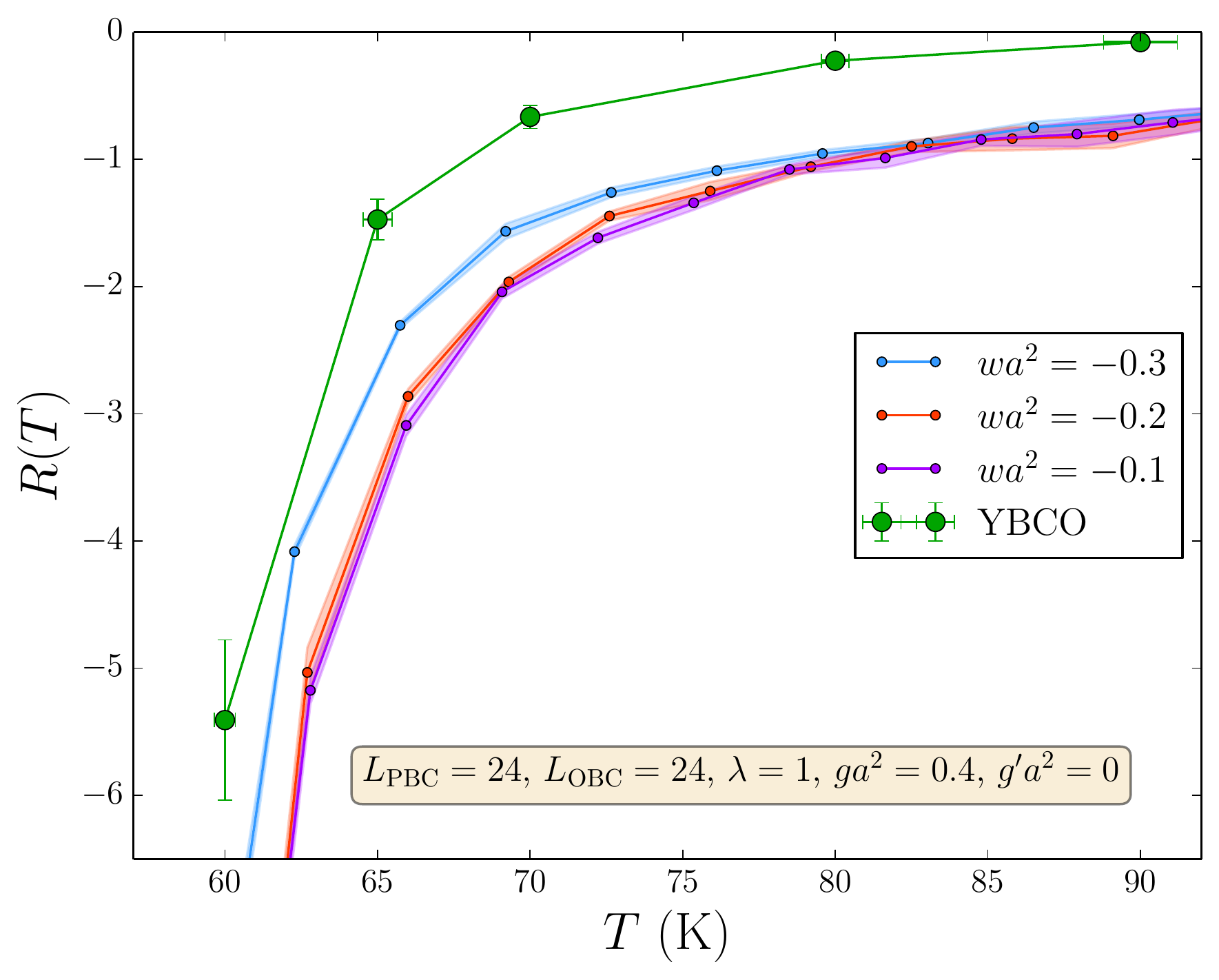}
         \caption{Comparison of $R(T)$ calculated in Monte Carlo simulations to data from X-ray scattering experiments\cite{achkar2} and torque magnetometry experiments\cite{ybcomag} on YBCO.  We illustrate results for the same parameter sets as in Figure~\ref{fig:SPhiVsT}. For each set of parameters $\lambda$, $ga^2$, $g'a^2$ and $wa^2$, the rescaling factor for the Monte Carlo data along the $T$-axis is determined from the $S_{\Phi_x}$ vs. $T$ fit. The shading for $R(T)$ accounts for statistical Monte Carlo errors as well as the uncertainty in the method for extracting $\xi_{\rm cdw}$ (see Figure~\ref{fig:corrLenVsT}).
         }
         \label{fig:ratioVsT}
\end{figure}

In Figure~\ref{fig:ratioVsT}, we compare our Monte Carlo simulations of the dimensionless ratio $R(T)$ (after determining $\rho_s$ as explained in Fig.~\ref{fig:SPhiVsT})
against the corresponding experimental values defined in Eq.~\eqref{Rexp} with no additional fitting parameters. 
The close correspondence between theory and experiment in both the absolute value and $T$ dependence of $R(T)$ 
is evidence that our model has captured significant aspects of the underlying physics. However, the theoretical values of $R(T)$ are consistently smaller than the experimental values; 
this discrepancy could be due to the different samples used for the diamagnetism and charge order measurements in computing
$R(T)$, or due to the limitations of our model, which are discussed in Section~\ref{sec:conc}.

\section{Conclusions}
\label{sec:conc}

This paper has presented Monte Carlo results on an effective classical model of competing superconducting and density wave orders
in the underdoped cuprates. Previous work \cite{science} has shown that the model can provide an excellent fit to the temperature dependence of the
structure factor of the density wave
correlations in the pseudogap regime, as measured by X-ray scattering experiments.
The present paper applied the same model to superconducting fluctuations as detected by diamagnetism measurements. We characterized the
strength of the diamagnetism by a dimensionless number $R(T)$, whose value is directly measurable in experiments, and which can also 
be conveniently computed in our Monte Carlo simulations. We found that the same set of fitting parameters used to describe X-ray scattering
also successfully capture the numerical value and variation with $T$ of $R(T)$.

However, the present classical model does omit some significant aspects of the physics. It does not include the effects of random field disorder 
acting on the charge order\cite{laimei}: we expect this to be important for enhancing the static component of the charge order at low $T$,
where there are deviations between our theory and the X-ray results. Interlayer couplings have also been omitted, and these will reduce the strength
of superconducting fluctuations above $T_c$, and possibly provide the needed correction to the theoretical value of $R(T)$. Our model also does not make
explicit reference to the fermionic degrees of freedom, but we believe these are properly accounted for by our effective
theory at the $T$ values of interest in the X-ray structure factor and the diamagnetism measurements.

With the improvements of our model just described, and precise experimental measurements of both diamagnetism and charge order correlations
on the same sample, we believe the prospects are bright for a precise quantitative theory of the pseudogap regime of the cuprate superconductors.

\acknowledgments
We thank D.~Chowdhury, J.~R.~Cooper, S.~Kivelson, M.~Gingras, Laimei Nie and L.~Taillefer for useful discussions.
This research was supported by the NSF under Grant DMR-1103860, the Natural Sciences and Engineering Research Council of Canada, the Perimeter Institute for
Theoretical Physics, the John Templeton Foundation, and the Canada Research Chair program.
Research at Perimeter Institute is supported by the
Government of Canada through Industry Canada and by the Province of
Ontario through the Ministry of Research and Innovation.

\FloatBarrier
\appendix

\section{Gaussian theory}
\label{app:gauss}

As alluded to in Section \ref{sec:model},
we tested our Monte Carlo method for calculating the linear-order diamagnetic susceptibility on a simple Gaussian theory for the SC order
\bea
H &=& \frac{\rho_s}{2} \sum_{\langle i j \rangle} |\Psi_i - \Psi_j|^2 + \frac{\rho_s}{2}  \sum_i \overline{\sigma} |\Psi_i|^2 \nn
&=&
\frac{\rho_s}{2} \left[ \sum_i (Z_i + \overline{\sigma}) |\Psi_i|^2 - \sum_{i,u} \Psi_{i}^\ast \Psi_{i+u} \right] \nn
&\equiv& \frac{\rho_s}{2} \sum_{i,j} \Psi_i^\ast \mathcal{M}_{i,j} \Psi_j , \label{Hgauss}
\eea
while ignoring all constraints, interactions, and the CDW components. Here $\mathcal{M}_{i,j}$ is a matrix defined by the expressions
above. This appendix drops factors of the inter-layer spacing $s$, and sets the lattice spacing $a=1$.
Then
\bea
\frac{M}{B T} &=& - \frac{1}{4L^2} \sum_{i,u}  \left(\epsilon_{\alpha\beta} r_{i \alpha} u_{\beta}\right)^2
\mathcal{M}^{-1}_{i,i+u} \label{gaussian} \\
&+& \frac{1}{4L^2 } \sum_{i,u} \sum_{j,u'}  \left(\epsilon_{\alpha\beta} r_{i \alpha} u_{\beta}\right)\left(\epsilon_{\gamma\delta} r_{j \gamma} u'_{\delta}\right) \mathcal{M}^{-1}_{i,j} \mathcal{M}^{-1}_{j+u',i+u}. \nonumber
\eea
Note that the right-hand-side is independent of $\rho_s$ and $T$: this is a special feature of the Gaussian theory.
The exact answer for the Gaussian theory in the limit $L \rightarrow \infty$ is\cite{science}
\bea
\frac{M}{B T} &=& - \int \frac{d^2 k}{4 \pi^2} \frac{8 \sin^2 (k_x) \sin^2 (k_y)}{(4 - 2 \cos(k_x) - 2 \cos(k_y) + \overline{\sigma})^4}
\label{exact} \\
&=& - \frac{1}{12 \pi \overline{\sigma}} \quad \mbox{as $\overline{\sigma} \rightarrow 0$}.
\eea
We compare the above expressions with our Monte Carlo results in Table~\ref{tab:gaussian}.
\begin{table}[htb]
\begin{tabular}{|c | c | c | c | c | }
\hline
$~~~\overline{\sigma}$~~~ & ~~~$L$~~~ & ~~$M/(BT)$~~ & ~~$M/(BT)$~~ & ~~Monte~~ \\
 & & Eq.~\eqref{gaussian} & Eq.~\eqref{exact} & ~~Carlo~~  \\
\hline
1 & 5   & -0.011490 &                           & -0.01149(2) \\
1 & 10 & -0.011206 &                           & -0.01124(5) \\
1 & 20 & -0.011108 &                           &  -0.0110(2) \\
1 & 40 & -0.011064 &                           &  -0.0113(6) \\
1 & 80 & -0.011043 &                           & -0.010(2)  \\ 
1 & $\infty$ & -0.011028(3) & -0.011024 & -0.0108(4) \\
\hline
0.5 & 5   & -0.038279 &                           & -0.03826(2) \\
0.5 & 10 & -0.034226 &                           & -0.03426(5) \\
0.5 & 20 & -0.032718 &                           & -0.0326(4) \\
0.5 & 40 & -0.032004 &                           & -0.031(1) \\
0.5 & 80 & -0.031656 &                           &  -0.026(4) \\
0.5 & $\infty$ & -0.03139(4) & -0.031315 & -0.0308(9)  \\
\hline
0.1 & 5   & -0.420271 &                          & -0.404(4)  \\
0.1 & 10 & -0.322547 &                          & -0.320(2) \\
0.1 & 20 & -0.268920 &                          &  -0.275(6) \\
0.1 & 40 & -0.247611 &                          &  -0.24(1) \\
0.1 & 80 & -0.237534 &                          &  -0.18(3)\\
0.1 &$\infty$ & -0.224(3) & -0.227827 &  -0.22(1) \\
\hline
\end{tabular}
\caption{Magnetization for the Gaussian theory $H$ in Eq.~\eqref{Hgauss}.
The extrapolation to $L=\infty$ in the third and fifth columns is performed by a least-squares fit to a quadratic polynomial of $1/L$, and the error bars for $L=\infty$ come from the covariance matrix of the least-squares fit. The Monte Carlo data in the fifth column was taken at $T=0.6$, although we also checked that the Monte Carlo results for $M/(BT)$ are independent of $T$.
}
\label{tab:gaussian}
\end{table}
The excellent agreement between the theory and the Monte Carlo results is strong evidence that our simulations have converged to the
thermodynamic diamagnetic susceptiblity.

\section{Superconducting phase}
\label{app:super}

For the superconducting phase we use the simple action
\beq
H_{sf} = \frac{\rho_s^R}{4} \sum_{ i,u} \left(\theta_i - \theta_{i+u} - A_{iu} \right)^2 + \frac{\rho_s^R m_\theta^2}{2} \sum_i \theta_i^2 ,
\eeq
where $\rho_s^R$ is the renormalized phase stiffness, and $m_\theta$ is a small mass added as an infrared regulator; the final result for the magnetization
will have a smooth limit as $m_\theta \rightarrow 0$.
So the current flowing along link $iu$ is
\beq
{\bf J}_{iu} = \mathbf{u} \rho_s^R \left(\theta_i - \theta_{i+u} - A_{iu} \right) .
\eeq
We can then write the total magnetic moment divided by the volume as
\beq
M = \frac{\rho_s^R}{4L^2 a^2 s} \sum_{i,u} \epsilon_{\alpha\beta} r_{i \alpha} u_{\beta} \left\langle
\theta_i - \theta_{i+u} - A_{iu} \right\rangle .
\eeq
Again, we expand $M$ to first order in $B$ and obtain
\bea
\frac{M}{B} &=& - \frac{\rho_s^R}{8 L^2 a^2 s} \sum_{i,u}  \left(\epsilon_{\alpha\beta} r_{i \alpha} u_{\beta}\right)^2  \nn
&+& \frac{(\rho_s^R)^2}{16T L^2 a^2 s} \sum_{i,u} \sum_{j,u'}  \left(\epsilon_{\alpha\beta} r_{i \alpha} u_{\beta}\right)\left(\epsilon_{\gamma\delta} r_{j \gamma} u^\prime_{\delta}\right) \nn
&~&~~~~~~~ \times \Bigl\langle  (\theta_i - \theta_{i+u})(\theta_j - \theta_{j+u'} ) \Bigr\rangle_0 ,
\eea
where the subscript indicates that this average is to be evaluated under $H_{sf}$ at zero field.
If we write the field-independent part of $H_{sf}$ as
\bea
H_{sf}^0 &=& \frac{\rho_s^R}{4} \sum_{ i,u} \left(\theta_i - \theta_{i+u}  \right)^2 + \frac{\rho_s^R m_\theta^2}{2} \sum_i \theta_i^2
\nn
&\equiv& \frac{\rho_s^R}{2} \sum_{i,j} \theta_i \mathcal{N}_{i,j} \theta_j ,
\eea
then
\bea
&&\frac{s M}{B \rho_s^R} = - \frac{1}{8 L^2 a^2} \sum_{i,u}  \left(\epsilon_{\alpha\beta} r_{i \alpha} u_{\beta}\right)^2  \nn
&&~~~~~~~+ \frac{1}{16 L^2 a^2} \sum_{i,u} \sum_{j,u'}  \left(\epsilon_{\alpha\beta} r_{i \alpha} u_{\beta}\right)\left(\epsilon_{\gamma\delta} r_{j \gamma} u^{\prime}_{\delta}\right) \nn
&&\times 
\left( \mathcal{N}_{i,j}^{-1} + \mathcal{N}_{i+u,j+u'}^{-1} - \mathcal{N}_{i+u,j}^{-1} - \mathcal{N}_{i,j+u'}^{-1} \right).
\label{latticesf}
\eea
We evaluate this expression numerically, and the $L \rightarrow \infty$ results are in 
precise agreement with the analytic results described below.

For an analytic expansion in the continuum limit, we can express the magnetization in terms of the $\theta$ propagator
$G_\theta (\mathbf{r}, \mathbf{r}')$. The propagator is conveniently expressed in terms of the eigenmodes of the Laplacian with 
Neumann (zero current) boundary conditions as
\beq
\frac{\rho_s^R}{T} G_\theta (\mathbf{r}, \mathbf{r}') = \sum_{m,n=0}^{\infty} \frac{\phi_{m,n} (\mathbf{r}) \phi_{m,n} (\mathbf{r}')}{ (m^2 + n^2) \pi^2/L^2 + m_\theta^2} ,
\eeq \\
{\parindent0pt where the eigenmodes are }
\beq
\phi_{m,n} (\mathbf{r}) = \left\{
\begin{array}{cc}
1/L,  & m=n=0 \\
(\sqrt{2}/L) \cos (m \pi (x/L - 1/2)), & n=0, m\neq 0 \\
(\sqrt{2}/L )\cos (n \pi (y/L - 1/2)), & m=0, n\neq 0 \\
(2/L )\cos (m \pi (x/L - 1/2)) & \\
~~~~\times \cos (n \pi (y/L - 1/2)), & n\neq0, m\neq 0 .
\end{array}
\right.
\eeq
Then, the continuum limit of Eq.~(\ref{latticesf}) at $m_\theta=0$ is 
\bea
\frac{s M}{B \rho_s^R a^2} &=& - \frac{1}{4 L^2} \int_{-L/2}^{L/2} dx \int_{-L/2}^{L/2} dy \, (x^2 + y^2) \nn
&+&  \frac{\rho_s^R}{4 T L^2} \int_{-L/2}^{L/2} dx \int_{-L/2}^{L/2} dy \int_{-L/2}^{L/2} dx' \int_{-L/2}^{L/2} dy' \nn
&~&~~~~~~~~~~
\Bigl[ (\mathbf{r} \times \nabla_{\mathbf{r}})(\mathbf{r}' \times \nabla_{\mathbf{r}'})G_\theta (\mathbf{r}, \mathbf{r}') \Bigr] \nn
&=& - \frac{L^2}{24} + \frac{L^2}{\pi^6} \sum_{m,n=1}^{\infty} (-1 + (-1)^m)^2 (-1 + (-1)^n)^2 \nn
&~&~~~~~~~~~~~~~~~~~~~~~~~~\times \frac{(m^2-n^2)^2}{m^4 n^4 (m^2 + n^2)} \nn
&=& -0.03514425 L^2  .
\eea
The numerical values of Eq.~(\ref{latticesf}) agree very well with the above result. The linear diamagnetic susceptibility of a two-dimensional superconductor in an $La \times La$ square geometry thus diverges 
as $-0.03514425 \, \rho_s^R (La)^2$ in the limit of large $L$;
this is, of course, a manifestation of the Meissner effect.

\end{document}